# Efficiency Discounted Exponential Growth (EDEG) Approach to Modeling the Power Progression of a Historical Dynasty

By Mark P. A. Ciotola, San Francisco State University, ciotola@sfsu.edu

30 October 2014

**Abstract.** Various approaches to model the progression of a dynasty in terms of power are discussed. The efficiency-discounted exponential growth (EDEG) approach is presented, and the effects of changing decay type and growth rate are demonstrated. The Russian Romanov dynasty is utilized as an example for several of the approaches. **Keywords:** dynasty, Romanov, exponential growth, efficiency-discounted exponential growth, EDEG

1.0 Introduction

This paper concerns generating models of the rise and fall of power of dynasties versus time. This sort of model can be called a power progression. Various approaches to modeling dynasties will be explored, then a new physical approach will be proposed. All of the approaches presented are simplifications that assume a gradual rise and fall. Of course history is rarely so cooperative. Hence, the models shown should be considered mere first approximations. Mathematical treatment of dynasties will be novel to most historians, so simple approaches will be discussed first, and more complex ones later.

It is simpler to model a sufficiently large, robust, independent dynasty than one that existed merely at the whim of its neighbors, for there are less significant dependencies, and thus can be approximated as a substantially isolated system. So we will utilize Russia's Romonov dynasty as an example. Widely accepted start and end dates are 1613 and 1917 (Mazour and Peoples 1975). Peter the Great and Catherine the Great were the two important rulers of the Romanov dynasty, and the Russian Empire gained much of its most valuable territory by the end of Catherine's reign. The Romanov dynasty was big, robust and essentially independent. It fought wars, but it generally was not under serious threat of extinction. Even Napoleon could not conquer Russia, but rather Russia nearly conquered Napoleon. This dynasty was reasonably long-lived, rather than just a quick, "flash-in-the-pan" empire.

1.1 Straight line

The simplest approach to model the power progression of a dynasty is a combination of two linear models. This requires start and end years of the dynasty, and its peak year as input parameters. For a single dynasty, the magnitude of the peak can be set to a nominal value of 1. Simply calculate (or draw) a straight line from the relative power value 0 from the start year of the dynasty to 1 at the peak year. How to select the peak year is less clear. A naïve approach would be to pick the chronological midpoint. An objective selection would be the date of maximum economic production, if that datum is available. Or one could choose a reasonable event, although this is less objective. For example, we will choose 1796, the year of the death of Catherine the Great, one of the most powerful rulers of Russia (Mazour and Peoples 1975). Figure 1 (left) shows this first line.

Next, calculate a straight line from magnitude 1 at the peak year to magnitude 0 at the end year, involving the slope and vertical-intercept, shown on Figure 1 (right). This model has several disadvantages among which it is discontinuous at its peak and assumes linear growth and decay. Further, it tells us little about what underlying causes and factors may be.

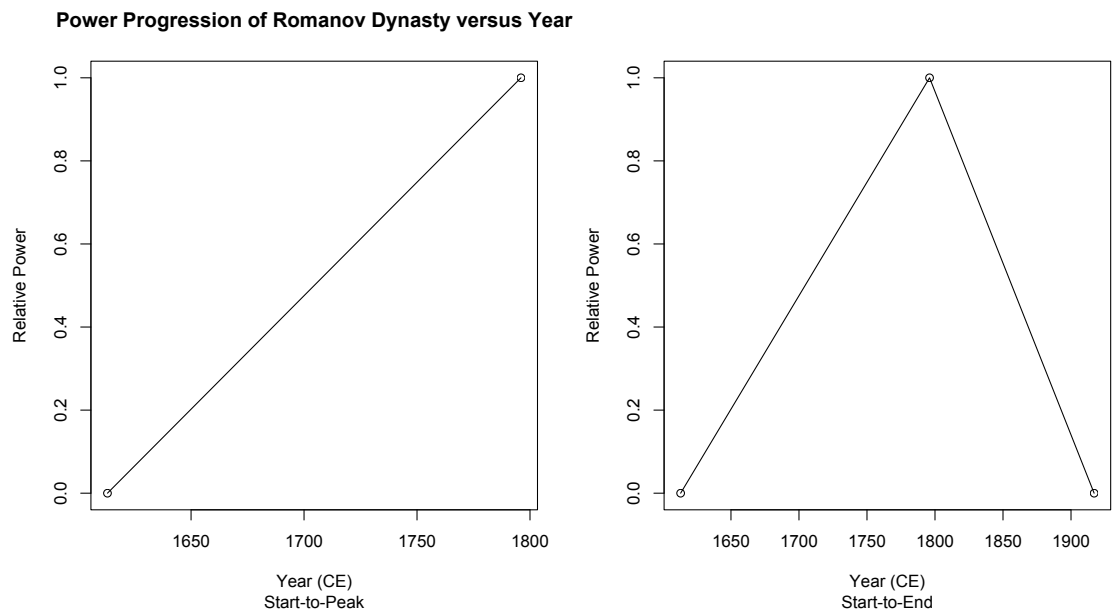

FIGURE 1: Line from start to peak year, where peak is death of Catherine the Great (left); Linear model of entire dynasty (right).

1.2 Inverted quadratic function

The next simplest approach is to use an inverted parabola function. Geologist M. King Hubbert considered this approach when he modeled peak oil (Hubbert 1980). As before, one can nominally set the peak magnitude. Here, one does not take the peak year as a parameter, but rather sets suitable parameters to that the parabola intersects the x-axis at the start and end dates of the dynasty (see Figure 2). This approach produces models with a rapid growth and decay yet with a relatively long period of relative stability in the midst. A chief disadvantage is that the symmetry of this function forces one to assume a peak year in the mid-point of the dynasty's life. Also, this model tells us very little about underlying causes.

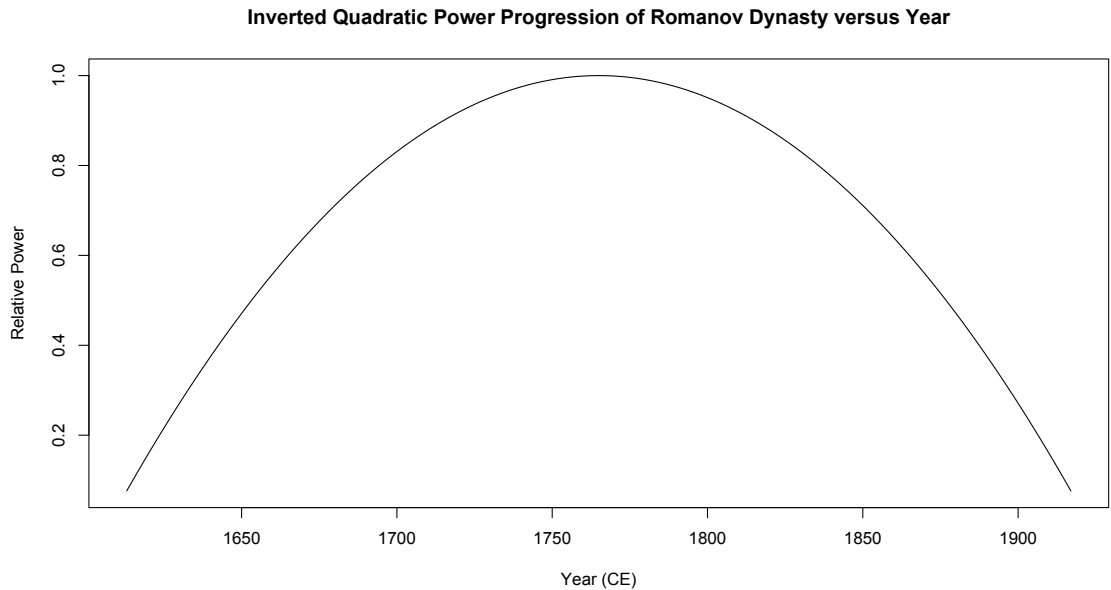

FIGURE 2: Inverted parabola with endpoints at 1613 and 1917.

1.3 Normal distribution

A more sophisticated approach is to model the dynasty as a normal distribution, or a mathematically similar function. Hubbert utilized bell-shaped plots for peak oil (Hubbert 1956, 1980). Such a function suggests that a dynasty comprises a collection of semi-random, resource-related events that cluster about the dynasty's peak. This approach can account for resource-based factors, where there is a degree of randomness in the ability to obtain such resources. For example, if there is a resource such as a critical mineral or petroleum where discoveries are to some degree by chance, this approach begins with few mining events, picks up with growth, then levels off with the difficulty of finding new deposits (Ciotola 1995). A chief disadvantage is that this model forces symmetry upon the dynasty's rise and fall. Another disadvantage is that one must literally begin with the peak and work to the endpoints. It provides no readily apparent means to *a priori* simulate the emergence of a dynasty.

The normal distribution approach also requires selecting a standard deviation value. Since the vertical axis represents power, and the area under the plot represents cumulative power, then one can set a standard deviation to produce the corresponding percentage of cumulative power. One can also determine the ratio of peak-to-start power, and use that to set the end-points. Note that a smaller standard deviation results in a sharper peak, whereas a greater one produces models that approach an inverted parabola (see Figure 3). This characteristic can be used to reject particular models based on available evidence.

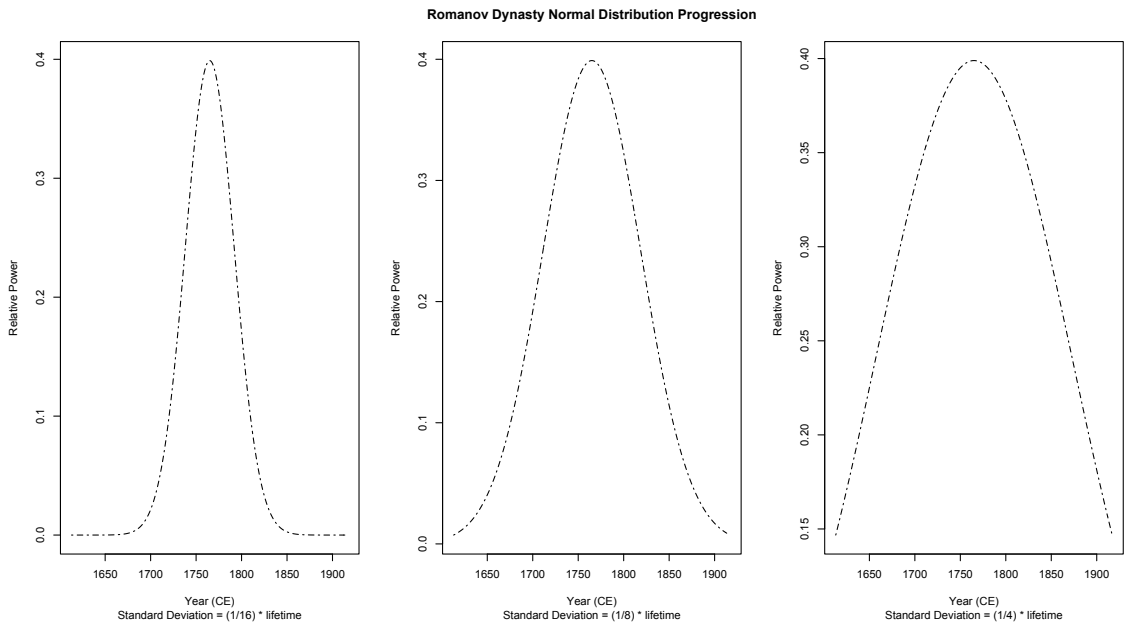

FIGURE 3: Normal distributions with range of standard deviations.

1.4 Maxwell-Boltzmann distribution

The Maxwell-Boltzmann distribution rises quickly then declines slowly. (It is possible to alter this distribution so that the opposite occurs). This distribution involves a quadratic function leading to a rise, and exponential decay leaving to a fall. So it is a qualitatively reasonable candidate to model the progression of a dynasty. The Maxwell-Boltzmann distribution has most of the advantages of the normal distribution, and it allows asymmetry. Disadvantages may include that this model is less simple both conceptually and mathematically, and requires more parameters than the approaches discussed above. An initial attempt to model the rise and fall (Ciotola 1995) of the Colorado San Juan mining region is shown in Figure 4. The parameters were adjusted to fit several data points provided by a historical account of the region by D. Smith (1982).

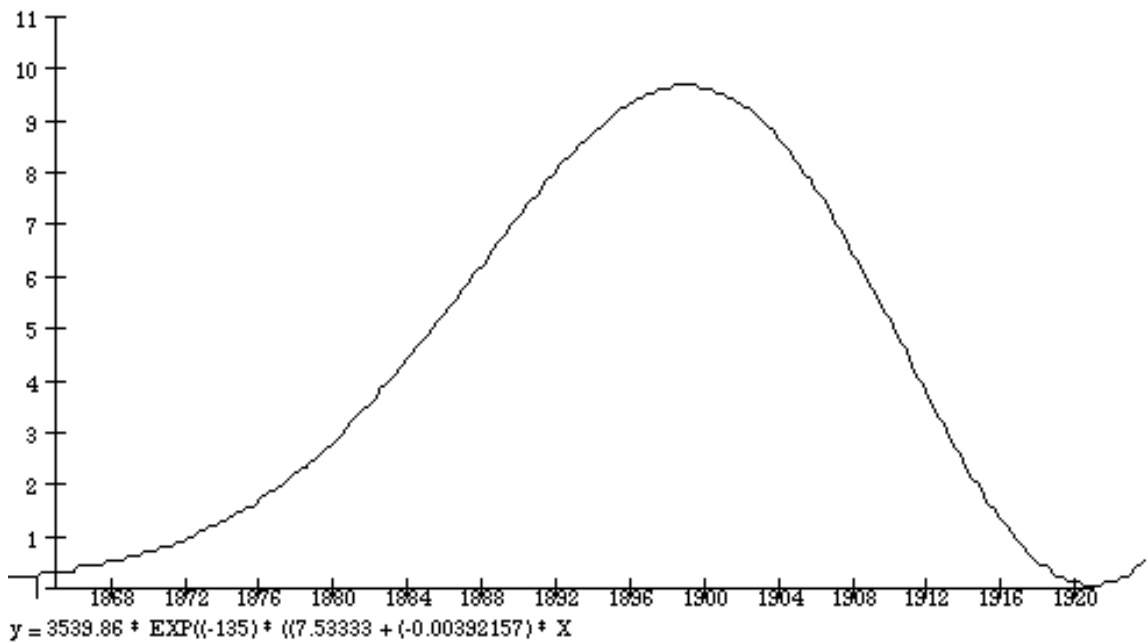

$y = 3539.86 * EXP((-135) * ((7.53333 + (-0.00392157) * X$

FIGURE 4: Maxwell-Boltxmann distribution for mining production versus year in San Juans region, with vertical axis represents millions of $US.

2.0 Efficiency Discounted Exponential Growth (EDEG) Approach

The efficiency-discounted exponential growth (EDEG) approach is relatively new. A few rough simulations were conducted earlier (Ciotola 2009, 2010), but this paper presents a more refined approach for generating dynasty models. EDEG produces a model based on two mathematically simple components, but allows the addition of other, more sophisticated components. By developing a fundamental approach to modeling the rise and fall of dynasties, it is possible to accept or reject models based upon both qualitative historical evidence and quantitative historical data.

Historical dynasties are consumers of energy and producers of power, so models in terms of such quantities are inherently fundamental in that they can be derived directly from the laws of physics and expressed in physical quantities. Such models are not *theories of everything*, but rather describe certain types of broad macro-historical phenomena rather than the intricate workings of the interactions of individual people.

The term *energy* is meant in the physical sense here. There are several possible measures of the physical energy of a dynasty, such as population governed or grain production. Each of these is translatable into physical units of energy. For example, the quantity of people multiplied by the mean Calorie diet per person will result in an amount in units of energy. These figures can be estimated for most dynasties over their lifespans, albeit with differing degrees on uncertainty. The proportion of this energy that rulers of the dynasty actually have at their disposal is beyond the scope of this paper, but should be considered for improved accuracy.

*Power* is a physical term. It refers to energy expended per unit of time. Yet it also has meaning within social and political contexts, and will be discussed in both senses. Absolute power would generally be presented in physical units of power such as Watts. However, it is possible to express *any* type of power in terms of proportions, such as the ratio of power at a dynasty's peak to its start date. Such a ratio can apply to physical, political or even military power. So the EDEG approach can be utilized to model any type of power. In fact, the EDEG approach provides a framework to explore the question of how political and physical power are related.

2.1 Origins of Efficiency Discounted Exponential Growth (EDEG)

This author's original efforts at modeling dynasties were heavily influenced by M. King Hubbert's attempts to model peak oil. The author originally used the normal distribution approach (Ciotola, 2001) discussed by Hubbert, but this approach was not sufficiently broad in that few historical dynasties utilized petroleum in major quantities, nor did Hubbert's attempts involve a driving tendency for history. The author began some related models for French and Spanish dynasties (2009) and more fully developed EDEG for petroleum modeling (Ciotola 2010).

2.2 Exponential Growth

It is noticed that systems in both nature and human society often grow exponentially (Ciotola 2001; Annila 2010). Exponential growth essentially means that a system's present growth is proportional to its present magnitude. For example, if a doubling of population (say of mice) is involved, then 10 mice will become 20 mice, while 20 mice will become 40 mice. A formula for exponential growth is:

$$y = e^{kt}$$

where *y* is the output, *k* is the growth rate, and *t* represents time.

Sources of growth can include geographic expansion, infrastructure improvements and trade expansion. It will be assumed that dynasties will strive to grow exponentially. (This paper does not attempt to prove this assertion, but rather it is a rebuttable presumption). If so, this certainly explains the rise of a dynasty. There is a minor distinction between exponential growth and compounded growth. Exponential growth essentially involves continuous compounding which produces a larger effective growth rate than discrete compounding. It is similar to the difference of quarterly versus daily compounding of a bank savings account. This effect is less significant at small growth rates but more so at very large rates. For the growth rates that we will consider, the effect is negligible compared to the other sources of uncertainty that exist.

The Romanov dynasty with various growth rates is shown in Figure 5. The plot shapes appear similar, except that a greater rate produces a "sharper" corner. Also, notice the range of power values: a greater growth rate produces a disproportionately greater power value at later points of time.

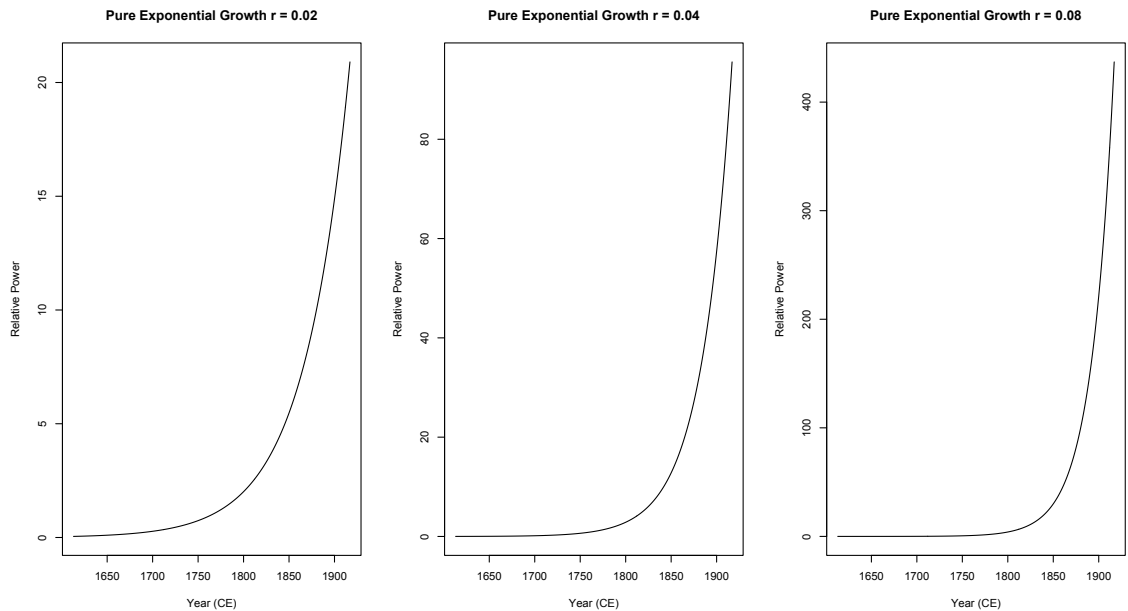

FIGURE 5: Plots of various pure exponential growth rates.

2.3 Decay

Dynasties inevitably end, which is typically preceded by a decline in power. Exponential growth alone is insufficient. Another physical principle comes to our aid. A dynasty can be viewed as a heat engine (or collection of such). Engines consume a potential to produce work or exert power. As an engine consumes a nonrenewable potential, the efficiency of that engine may decrease. (For physics-savvy readers, picture a Carnot engine operating across exhaustible thermal reservoirs. As heat is transferred, the temperature difference will decrease, and so to will efficiency (Ciotola, 2003)). Therefore the engine's net production will decrease and eventually fall to zero.

Likewise, as the dynasty progresses, non-renewable resources will be consumed, and efficiency will decrease. There will still be production until the end, but there will be a lower *return on investment*, so to speak. Causes of decay can include overuse of agricultural land leading to nutrient depletion, the build-up of toxins in the environment, depletion of old growth forests, and even running low on social goodwill.

There are two types of decay, linear and exponential, compared in the plot below. Figure 6 contains plots of linear and exponential decay. Note that efficiency is shown as a multiplier rather than a percentage.

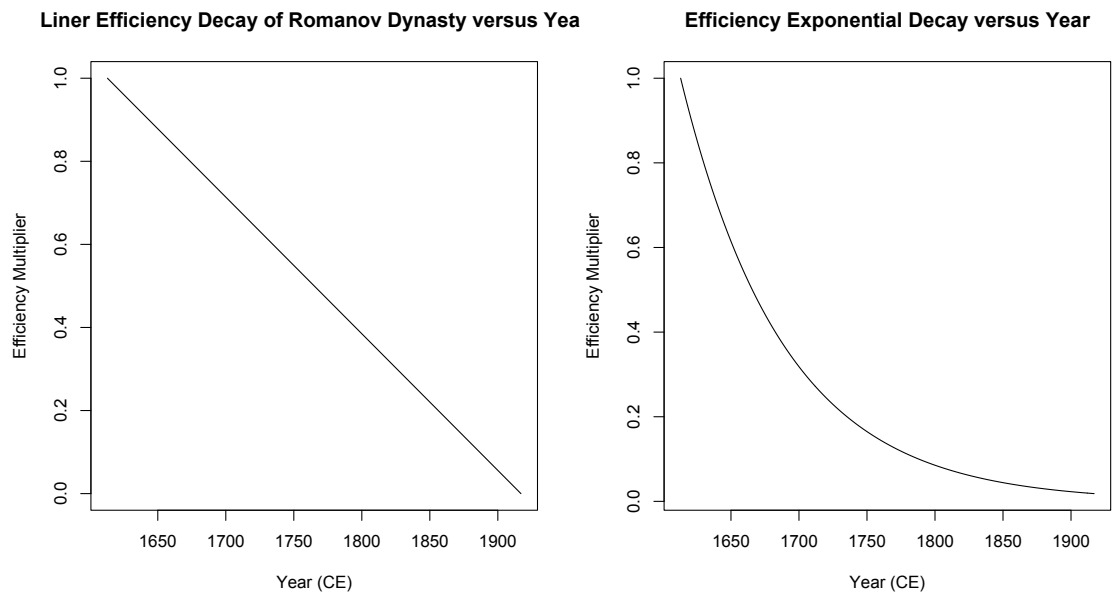

FIGURE 6: Plots of linear and exponential decay.

Exponential decay of efficiency is produced by a Carnot engine operating across an exhaustible thermodynamic potential. It is a fundamental form of decay. The following is an equation for an exponential decay function:

$$y = e^{-kt},$$

where *y* is the output, *k* is a constant of proportionality, and *t* represents time. This appears nearly identical to the exponential growth function, except that the exponent is negative. However, there are two disadvantages of exponential decay within the context of modeling dynasties. First, it is slightly difficult to set up. For example, exponential decay has an infinitely long tail. While this allows for mathematical immortality, most of the tail is superfluous in the context of a dynasty of limited lifetime. Second, it may not provide the most consistent models with observations.

A linear approach is simpler to set up. Importantly, it also provides some reflection of efficiencies achieved through centralization and economies of scale as the dynasty progresses. It has unambiguous beginning and end points. Efficiency cannot be greater than one, and is typically no lower than zero. Therefore, as a first approximation, one can set the efficiency to 100% at the start date of the dynasty and 0% at the end year (except that the math is simpler if the value 1 is used for 100%). Using a value of 0 for ending efficiency ensures that the dynasty actually does end by its historical end date. Although physical efficiency is typically lower than 100% for real life heat engines, 1 provides an easy starting point that also produces the correct shape of curve. The following is an example of linear decay function:

efficiency = 1 – ((year – start year)/(end year – start year)).

As the year increases, efficiency will decrease. Using a lower initial efficiency reduces the magnitude of production increase for the dynasty compared to its initial production. It also flattens out the curve. Using a value of zero for ending efficiency ensures that the dynasty actually does end by its historical end date. It is possible to use a value other than zero for the ending efficiency, but then some other factor must be used to end the dynasty.

2.4 Efficiency-Discounted Exponential Growth (EDEG)

We still need to go a step further. So now we bring exponential growth and declining efficiency together. We need to use the decay to discount exponential growth, just a little in the beginning, then completely at the end. The following is an example of an EDEG equation:

$$y = \text{exponential growth function} * \text{efficiency function},$$

where * is a multiplication symbol. Substituting in our functions (utilizing linear decay):

$$Y = e^{kt} *(1 - (t/(\text{end year} - \text{start year}))).$$

$$Y = e^{k(\text{year} - \text{start year})} *(1 - ((\text{year} - \text{start year})/(\text{end year} - \text{start year}))).$$

This produces a steady rise, a level period and a slightly faster decay. So by discounting exponential growth by decreasing efficiency, we then have a rise and fall pattern that is consistent with the rise and fall of a dynasty.

2.5 Application

Let us apply the EDEG approach to the Romanov dynasty. Let us assume a conservative 1% growth rate. Let us further assume linear decay from 100% to 0% efficiency. A simulation has been written in the Ruby programming language. This language is mathematically robust, yet involves code that is relatively easy to read and understand. The dynasty is run through the Ruby simulator, using the above parameters. A data table was produced to generate in the plot shown in Figure 7.

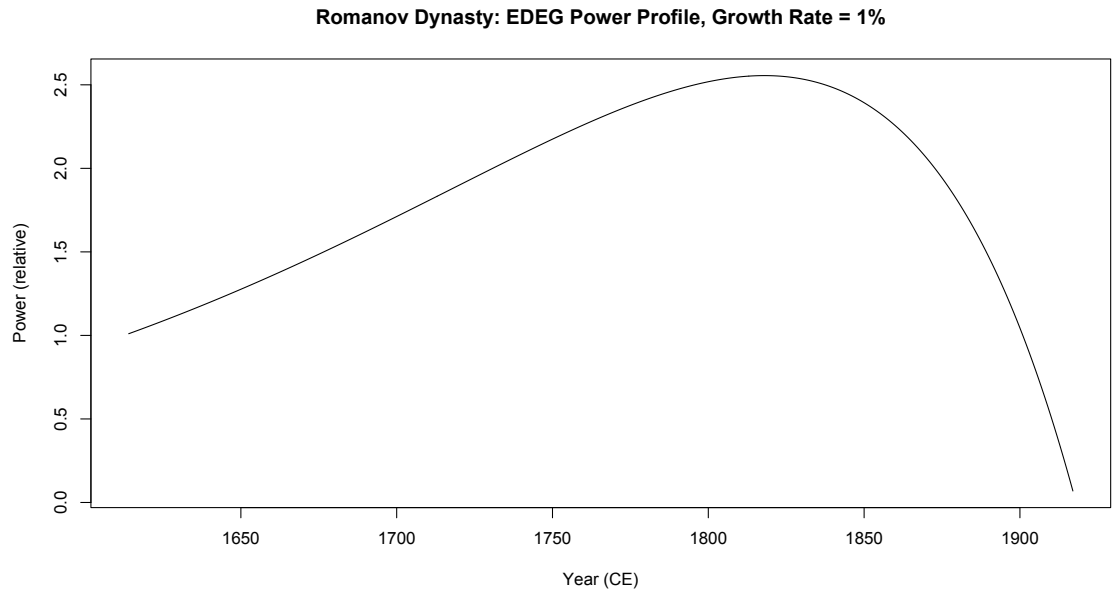

FIGURE 7: EDEG power progression for Romanov dynasty with growth rate of 1%.

Here the peak is close to 1820. Napoleon had been conquered, and the dynasty had achieved much of its geographic expansion by them. Yet by this time, social unrest began to shake the Romanov dynasty. Also, note how the dynasty power begins at a level of 1 and ends at a level of 0. This is appropriate, since the dynasty had to begin from something, but typically ends in nothing. For example, the ancestors of the Romanovs existed before 1613, but the entire immediate family was killed during the Russian revolution. The peak occurs at a relative power value of height of 2.6, which indicates that the dynasty was over twice as powerful at its peak as it its beginning. Remember, this model is merely a hypothesis that is either valid or invalid for a particular level of uncertainty.

The simulation was run again with higher growth rates. We again assume linear decay from 100% to 0% efficiency (see Figure 8). Note several in the response of power to a changing growth rate parameter. First, a higher growth rate results in a later peak. Second, the total peak to initial power ratio skyrockets as the growth rate is increased.

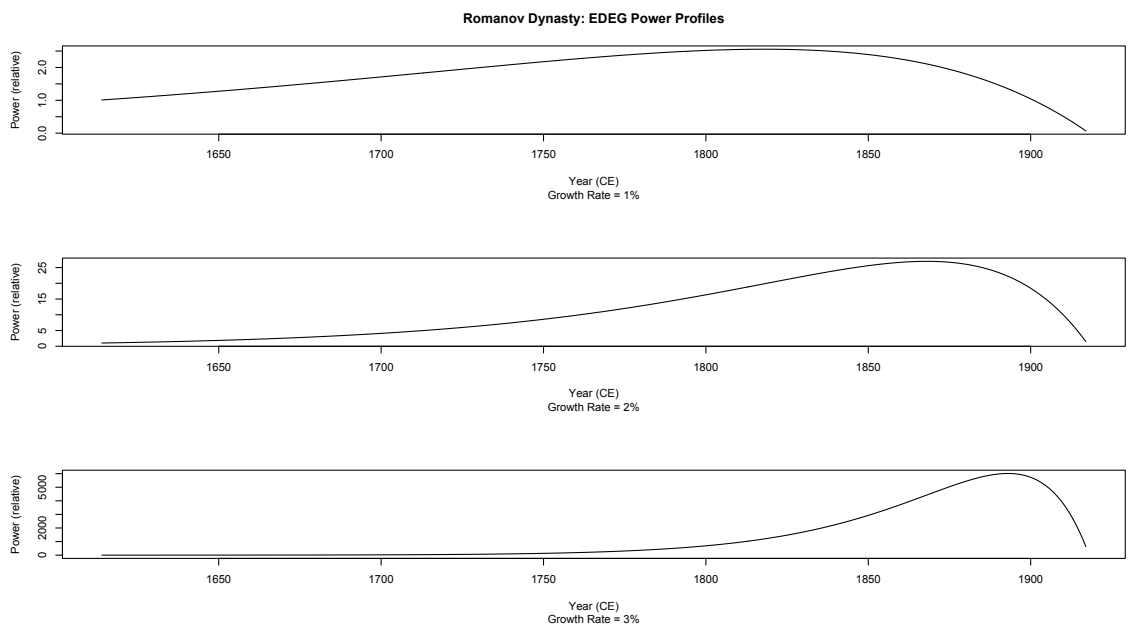

FIGURE 8: EDEG power progression for Romanov Dynasty for various growth rates.

Additional factors can be imposed as adjustment functions. One-time events (such as a rare but large natural disaster) can be superimposed as an event "mask". It may be of further interest to tie the rise and fall to patterns concerning the production and consumption of resources, to determine what correspondence, if any, there is between physical and political power. This can be explored by utilizing actual physical energy data to produce a model of physical power, and then comparing that model with evidence of political power versus time. With the wealth of historical data being gathered in anthropological data warehouses, and other "big data" facilities, this may be accomplished with increasing validity.

3.0 Discussion and Future Directions

This presentation of the EDEG approach is more of a barebones beginning than a complete end. It raises more questions than it answers, but it enables a broad framework to answer these questions. This framework acts as a unifying skeleton to link the humanistic elements of history with the quantitative constraints of the physical universe.

  The power of such a framework should not be underestimated. It is possible to gather quantitative data (or quantify qualitative evidence), perform statistical analysis and accept or disprove hypotheses. Yet such results, while often important, are merely empirical. They are often hard to use to constrain or illustrate each other. In a unified framework, all results act to constrain all other results. When we learn about one thing, we necessarily learn something about everything else. This is where the physical sciences have derived much of their strength.

  There are many immediately apparent improvements to improve the value of the EDEG approach. One improvement would be to better understand efficiency decay. Another improvement would be to start using actual data of physical energy, to the extent such data is available. Another improvement will be to separate the power level of the underlying society from that of the dynasty. For example, Russia did not disappear upon the death of the Romanov dynasty. On the contrary, it is still one of the most powerful societies on Earth. The brings up the need to be able to model the emergence of a series of dynasties in a way that connects and constrains each dynasty, such as concerning relative strength and timing of emergence. Further, there needs to be a way to compare co-existing dynasties and model their interaction within this framework. While the EDEG approach suggests possible means, the devil will be in the details.